\documentclass[]{aa}
\usepackage{sidecap}
\usepackage{graphicx}
\usepackage[varg]{txfonts}
\usepackage{url}

\usepackage{lscape}
\usepackage[dvipsnames]{xcolor}
\usepackage{hyperref}
\begin{document} 
\title{The mid-infrared environment of the stellar merger remnant V838\,Monocerotis}
   \author{Muhammad Zain Mobeen \inst{1}
          \and
          Tomasz Kami{\'n}ski\inst{1}
          \and
          Alexis Matter\inst{2}
          \and
          Markus Wittkowski\inst{3}
          \and 
         Claudia Paladini\inst{4}
          }

\institute{\centering
Nicolaus Copernicus Astronomical Center, Polish Academy of Sciences, Rabia{\'n}ska 8, 87-100 Toru{\'n}, Poland 
\and
Université Côte d’Azur, Observatoire de la Côte d’Azur, CNRS, Laboratoire Lagrange, France  \label{inst2}
\and
European Southern Observatory, Karl-Schwarzschild-Str. 2, 85748 Garching bei Munchen, Germany\label{inst3}
\and 
European Southern Observatory, Alonso de Cordoba 3107, Vitacura, Santiago, Chile\label{inst4}
}
\authorrunning{Mobeen et al.}
\titlerunning{MATISSE observations of V838 Mon}
 
\abstract{In 2002, V838 Monocerotis (V838\,Mon) erupted in a red novae event which has been interpreted to be a stellar merger. Soon after reaching peak luminosity it began to cool and its spectrum evolved to later spectral types. Dust was also formed in the post-merger remnant making it bright in the mid-infrared. Interferometric studies at these wavelengths have suggested the presence of a flattened elongated structure.}{ We investigate, for the first time, the structure and orientation of the dusty envelope surrounding V838\,Mon in the $L$(2.8--4.2 $\mu$m) band using recent observations with the MATISSE instrument at the VLTI.}{We perform simple geometrical modelling of the interferometric observables using basic models (disks, Gaussians, point sources, along with their combinations). We also reconstructed an image and analyzed the corresponding $L$-band spectrum.}{This study indicates the presence of an elongated, disk-like structure near 3.5\,$\mu$m, similar to what has been observed in other wavelength regimes. In particular, the orientation at a position angle of --40\degr\ agrees with prior measurements in other bands.}{The dusty elongated structure surrounding V838 Mon appears to be a stable and long lived feature that has been present in the system for over a decade. Its substructure and origin remain unclear, but may be related to mass loss phenomena that took place in the orbital plane of the merged binary.}

\maketitle

\section{Introduction}\label{intro}
In early 2002, V838\,Monocerotis (V838\,Mon) erupted in a red nova event \citep{2002A&A...389L..51M,2005A&A...436.1009T}. On time scales of a few weeks it increased its luminosity by over two orders of magnitudes, reaching a peak luminosity of $10^6$\,L$_{\sun}$ \citep{2005A&A...436.1009T,2008AJ....135..605S}. The main mechanism behind this event is thought to be the merger of an 8\,M$_{\sun}$ main sequence star with a 0.4\,M$_{\sun}$ young stellar object \citep{2006A&A...451..223T}. Soon after the eruption, the temperature of V838\,Mon decreased and its spectra evolved to an L-type supergiant \citep{2003MNRAS.343.1054E, 2015AJ....149...17L}. Spectral studies revealed the presence of a myriad of circumstellar molecules, including water and transition-metals oxides \citep{2002A&A...395..161B,2009ApJS..182...33K}. Dust formation also took place in the merger ejecta \citep{2008ApJ...683L.171W,alma}. Furthermore, a B-type companion was discovered close to V838\,Mon \citep{Bdiscovery2} which suggested that the merger occurred within a triple system. The companion disappeared in 2005 in the optical after becoming shrouded by the dusty ejecta produced in the 2002 eruption \citep{2009A&A...503..899T}. V838\,Mon remains a prototype of a luminous Galactic red nova, a type of transient whose analogs have been observed in the Local Group \citep{2019A&A...630A..75P}. 

The stellar remnant of V838\,Mon and its dusty environment have remained sufficiently bright for mid-infrared (MIR) interferometric measurements. The first such observations were done in late 2004 by \cite{2005ApJ...622L.137L}. They obtained $K$ band (2.2\,$\mu$m) measurements of squared visibilities using the Palomar Testbed Interferometer (PTI). They were also able to constrain the angular diameter of the central star of $1.83 \pm 0.06$\,mas. The measurements suggested an asymmetric source, but they were too sparse to define the geometry of the object. Followup interferometric observations were carried out by \cite{2014A&A...569L...3C}, between 2011 and 2014, using two Very Large Telescope Interferometer (VLTI) instruments: Astronomical Multi-BEam combineR \citep[AMBER;][]{2007A&A...464....1P} in $H$ and $K$ bands, and MID-infrared Interferometric instrument \citep[MIDI;][]{2003Ap&SS.286...73L} in $N$ band. Uniform disk fits to the AMBER data yielded a stellar diameter of $1.15 \pm 0.2$\,mas. Based on this, the authors claimed that the photosphere had shrunk by about 40\% in the span of a decade. Furthermore, their AMBER data modeling suggested an extended component (in addition to the star) with a lower limit on the full width at half-maximum (FWHM) of $\sim$20\,mas. MIDI data modeling suggested further that V838\,Mon is surrounded by an extended flattened structure whose major axis length changes as a function of wavelength between 25 and 70\,mas in $N$ band. More recent interferometric observations at millimeter (mm) wavelengths with the Atacama Large Millimeter/submillimeter Array (ALMA) revealed the presence of a flattened structure with a FWHM of 17.6$\times$7.6\,mas in the immediate surroundings of V838\,Mon \citep{alma}. The multi-wavelength measurements, from 2.2\,$\mu$m to 1.3\,mm, represent mainly dust emission and they all seem to be similarly oriented, with the major axes at position angles (PAs) of --10\degr\ (MIDI) to --50\degr\ (ALMA). The observed circumstellar features could be manifestations of a single structure, such as a disk seen at an intermediate inclination. The presence of such a disk is expected in merger remnants as a reservoir of angular momentum that was previously stored in the merged binary system \citep[e.g.][]{2017ApJ...850...59P}.    

The stellar remnant of V838\,Mon and its circumstellar environment are of prime interest because they illustrate directly what happens to a system years to decades after the stellar coalescence. They provide us with crucial information on the physical processes governing mergers of non-compact stars and the recovery phase of the merger product to a stable configuration. Here, we report recent MIR interferometric observations with the VLTI that constrain the structure of the closest vicinity of the merger product 18 years after the red nova eruption and 6 years after the last VLTI observations.   

\section{Observations and data reduction}
V838\,Mon was observed with the Multi AperTure mid-Infrared SpectroScopic Experiment (MATISSE) instrument \citep{2014Msngr.157....5L}, which is a four-beam interferometer at the VLTI. MATISSE observes simultaneously in the $L$ (2.8--4.2 $\mu$m), $M$ (4.5--5 $\mu$m), and $N$ (8--13 $\mu$m) bands, but due to technical limitations in 2020 the observations used in this study were taken in the $L$ and $N$ bands. The source was observed in January and March 2020 at seeing conditions indicated in Table\,\ref{table:Table 1}. V838\,Mon and a calibrator HD\,52666 were observed in the SCI-CAL sequence with two extended configurations (Table\,\ref{table:Table 1}) of the 1.8\,m Auxiliary Telescopes (ATs) and at a low spectral resolution (R$\sim$30). On 20 January, the calibrator star could not be observed due to technical issues and the night is excluded from this study.

\begin{table}
\caption{V838\,Mon observation log.}
\centering
\begin{tabular}{c c c c}
\hline
Date & Configuration & Seeing & Quality \\ 
\hline
01/03/2020&A0-G1-J2-K0&0\farcs66 &Low visibilities \\
01/19/2020&A0-G1-J2-K0&0\farcs43 &Good \\
01/20/2020&A0-G1-J2-K0&0\farcs40&No calibration \\
03/01/2020&A0-G1-J2-J3&0\farcs59&Good \\ 
\hline
\end{tabular}
\label{table:Table 1}
\end{table}

\begin{table}
\caption{Total $LMN$-band fluxes of V838\,Mon and the calibrator.}
\centering
\begin{tabular}{c c c c c}
\hline
Object & $L$ band & $M$ band & $N$ band & Reference\\  
&[Jy]&[Jy]&[Jy]  \\
\hline
V838\,Mon&5&4&30 &1\\
HD\,52666&132&67&18 &2 \\
\hline
\end{tabular}
\tablebib{(1)~\cite{alma}; (2) \cite[MDFC;][]{2019yCat.2361....0C}}
\label{table:Table 2}
\end{table}

A MATISSE observation sequence consists of a series of twelve 1-min exposures where simultaneous interferometric and photometric data are acquired in $L$-band, while four interferometric exposures followed by eight photometric exposures are acquired in $N$-band. The first series of four exposures is taken without chopping, while the following eight exposures are executed with chopping between the target and the sky. Each exposure is taken in one of four configurations of two beam commuting devices (BCD), which switch pairwise the telescopes beams. This process helps in removing instrumental systematic effects. 

We used the MATISSE pipeline version 1.5.5\footnote{\url{https://www.eso.org/sci/software/pipelines/matisse/}} to reduce our raw data. The resulting OIFITS files (version 2) contain uncalibrated interferometric observables, including six dispersed squared visibilities and three independent closure phase measurements per exposure.


The observations of V838\,Mon were then calibrated using our measurements of the K5\,III star, HD\,52666, whose characteristics were  taken from the Mid-infrared stellar Diameters and Fluxes compilation Catalogue\footnote{MDFC is available through VizieR service at https://vizier.u-strasbg.fr/viz-bin/VizieR} \citep[MDFC;][see Table\,\ref{table:Table 2}]{2019yCat.2361....0C}. In particular, the visibility calibration was performed by dividing the raw squared visibilities of V838\,Mon by these of the calibrator, corrected for its diameter (known as the interferometric transfer function). The calibrated total spectra of V838\,Mon were obtained by multiplying the ratio between the target and calibrator raw fluxes measured by MATISSE at each wavelength, followed by multiplication by a model of the absolute flux of the calibrator. This model was taken from the PHOENIX stellar spectra grid as described in \cite{2013A&A...553A...6H}.  

\begin{figure}
    \centering
    \includegraphics[width=\columnwidth]{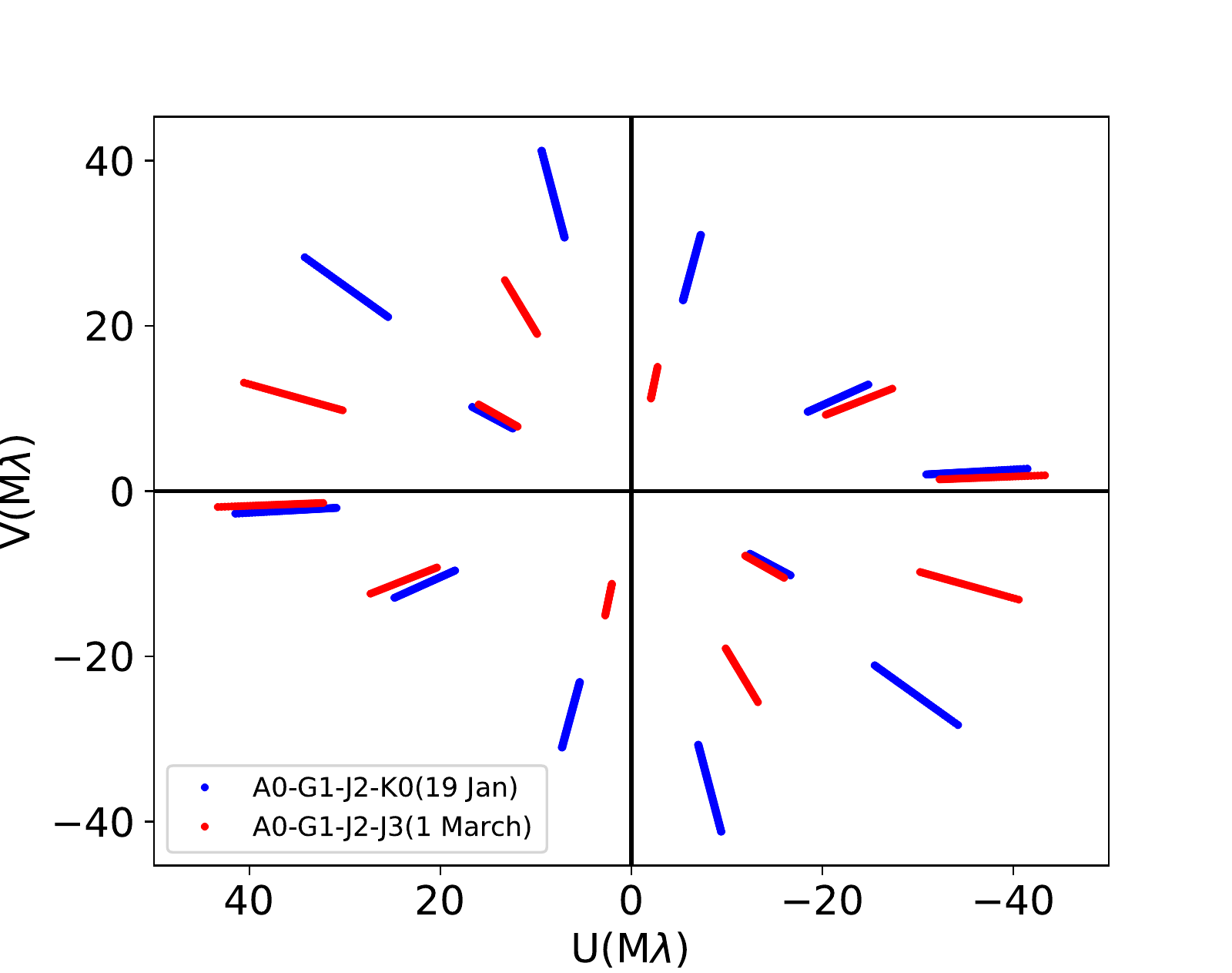}
    \caption{$UV$ plane tracks for 19 January and 1 March in $L$ band.} 
    \label{Fig.1}
\end{figure}

The calibration was done for data in the $L$ and $N$ bands. Unfortunately, our $N$ band observations of V838\,Mon turned out to be of insufficient quality due to too low source brightness at observed baselines. Indeed, the $N$-band correlated flux is below 5\,Jy at all baselines, which is at the sensitivity limit for MATISSE with the ATs and is much below the expected total-power flux \citep[$\sim$30\,Jy;][]{alma} of the source. The data were thus excluded from our analysis. We also visually inspected the $L$ band squared visibilities and closure phases using the OIFITS viewer by Jean-Marie Mariotti Center\footnote{\url{https://www.jmmc.fr/english/tools/data-analysis/oifits-explorer/}} (JMMC). For the night of 3 January 2020, the measured squared visibilities were almost two orders of magnitudes lower than those from the other nights. We compared the $UV$ coverage of the two nights and found that they did not vary by much. Therefore, we concluded that the observations on January 3 were faulty and excluded them from the analysis as well. That left us with two observations. The $UV$ coverages for the two observations in $L$ band are shown in Fig.\,\ref{Fig.1}.

Given the relatively low $L$-band total flux of V838\,Mon of about 5\,Jy, only the chopped exposures provide reliable photometry and thus reliable absolute visibility measurements. We thus considered only the chopped exposures for the squared visibilities. We then again visually inspected the squared visibilities for each of the chopped exposures in the different BCD configurations\footnote{\url{https://www.eso.org/sci/facilities/paranal/instruments/matisse/doc/MATISSE_USERMANUAL.pdf}}. This was done with the help of python scripts written by the MATISSE Consortium\footnote{\url{https://gitlab.oca.eu/MATISSE/tools/-/tree/master}}. For 19 January, we found outlier values off by more than 3$\sigma$ in the BCD in-in exposures. We discarded these exposures and considered only the ones with the BCD out-out configuration. 
For our analysis, we further restricted the data to the wavelength range 2.95--4.00\,$\mu$m to exclude data points affected by telluric absorption. The calibrated squared visibilities and closure phases are shown in Fig.\,\ref{Fig-A1}. The error bars shown there contain the short-term error affecting the individual 1-min exposures, as computed by the pipeline, and the standard deviation between the merged exposures. A calibration error associated with the stability of the $L$-band transfer function over the night is of about 2\% (Lopez et al., in prep.) and was neglected in our error budget.   

The nominal field of view (FoV) of our observations is $\approx$600\,mas. For the purpose of estimating roughly the maximum angular resolution, we use the Rayleigh criterion, $\theta\!\approx\!\lambda / B$, where $B$=140\,m is the size of the longest baseline. For the relevant wavelength range, we get resolutions of about 4--6\,mas.

\section{Analysis}
\subsection{$L$-band spectrum}\label{sec-spec}
This is the first time that interferometric observations of V838\,Mon are presented in the $L$ band. This range of the spectral energy distribution (SED) of V838\,Mon is dominated by thermal emission of circumstellar dust, with negligible contribution of flux coming directly from the stellar photosphere \citep{alma,woodward}. Here, we attempt to identify the dominant spectral components in the band. 

\begin{figure}
    \centering
    \includegraphics[width=\columnwidth]{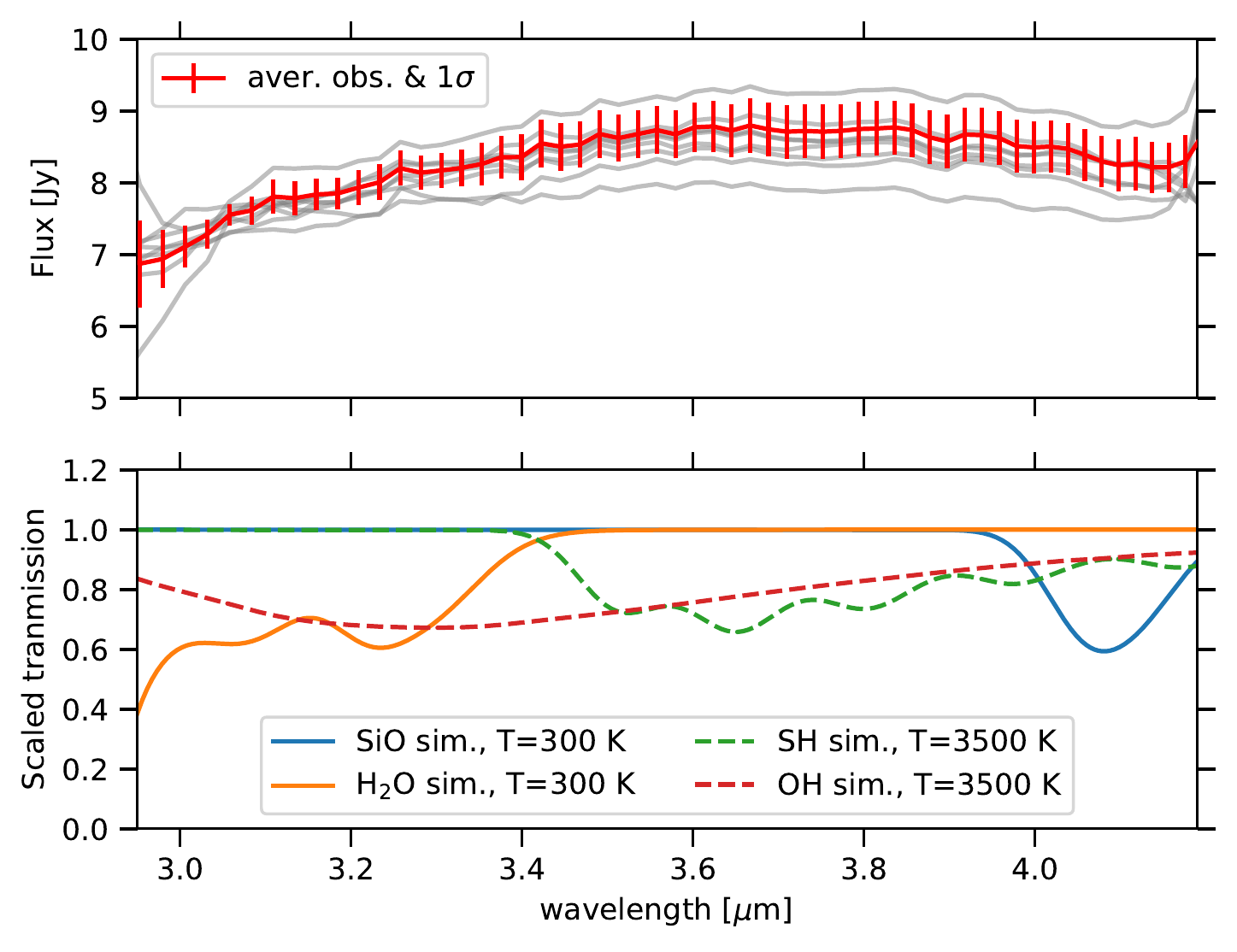}
    \caption{$L$ band spectra of V838\,Mon. The solid gray lines represent individual calibrated spectra for both nights. The red line is a weighted-mean average spectrum, with errorbars representing one standard deviation in individual spectra. The bottom panel shows opacity curves for four species. They are arbitrarily scaled. All simulations were smoothed to match the resolution of the observations. SiO and H$_2$O curves were simulated for circumstellar gas temperatures, while SH and OH were assumed to form in photospheric temperatures. The latter two are unlikely to be present in the spectrum of V838\,Mon.}
    \label{fig-spec}
\end{figure}

The individual flux-calibrated spectra and a weighted mean spectrum from both nights are presented in Fig.\,\ref{fig-spec}. Based on MIR studies of total-power spectra of the object from 2003, our spectrum should be affected by circumstellar features of SiO and H$_2$O, and may contain contributions of photospheric bands of OH and SH \citep{lynch}. We obtained opacity curves for all four molecules in {\tt pgopher}\footnote{\url{http://pgopher.chm.bris.ac.uk}} \citep{pgopher}. Molecular spectroscopic data were extracted from ExoMol\footnote{\url{https://www.exomol.com}} \citep{exomol} and are based on \cite{water} (H$_2$O), \cite{sio} (SiO), \cite{sh} (SH), and \cite{oh} (OH). The opacity curves are presented in Fig.\,\ref{fig-spec}. The simulations illustrate band shapes in the optically thin limit. Scaling of the bands was arbitrary, as we were only interested in band identification. 

Our simulations show that the spectrum can be explained by a combination of absorption of warm water in the 2$\nu_2$ band, dominant below 3.4\,$\mu$m, and of SiO in its first-overtone ($\Delta \nu$=1) band at around 4.1\,$\mu$m. The shape of the water band suggests an excitation temperature of about 200--700\,K, typical for circumstellar absorption in this source. Note that in 2003, shortly after the outburst, SiO $\Delta \nu$=1 was in emission. At the resolution of the observations, the photospheric bands of SH and OH, considered to be present in 2003 by \cite{lynch}, would not be discernible in the observed spectrum. In our simulations, we assumed they form at a photospheric temperature of 3500\,K \cite[cf.][]{alma}. Indeed, in the SED model of \citet{alma} the photospheric fluxes are negligible in the $L$ band. 

We conclude that a big part of the $L$-band flux is circumstellar dust emission that is not affected strongly by circumstellar absorption. The absorption is strongest near $\approx$3.0\,$\mu$m but even there it is not stronger than 20\% of the continuum. 

\subsection{Geometric models}\label{sec-litpro}
Due to the sparse number of points in the $UV$ plane, we mainly analyzed the data by comparing the squared visibilities ($V^2$) and closure phases to simple geometric models of the source. The analysis was performed using LITpro\footnote{\url{https://www.jmmc.fr/english/tools/data-analysis/litpro/}} \citep{2008SPIE.7013E..44T}. LITpro allows the user to model observables with source shapes such as a point source, a circular disk, a circular Gaussian, or their combinations. LITpro uses a $\chi^2$ minimization to find the best model and provides the reduced chi-squared, $\chi^2_r$, on the output. We simultaneously modeled the calibrated squared visibilities and closure phases. These quantities remain fairly constant with wavelengths, except at the edges of the spectral band, where the errors are very high. 

\begin{figure*}
    \centering\sidecaption
    \includegraphics[width=12cm]{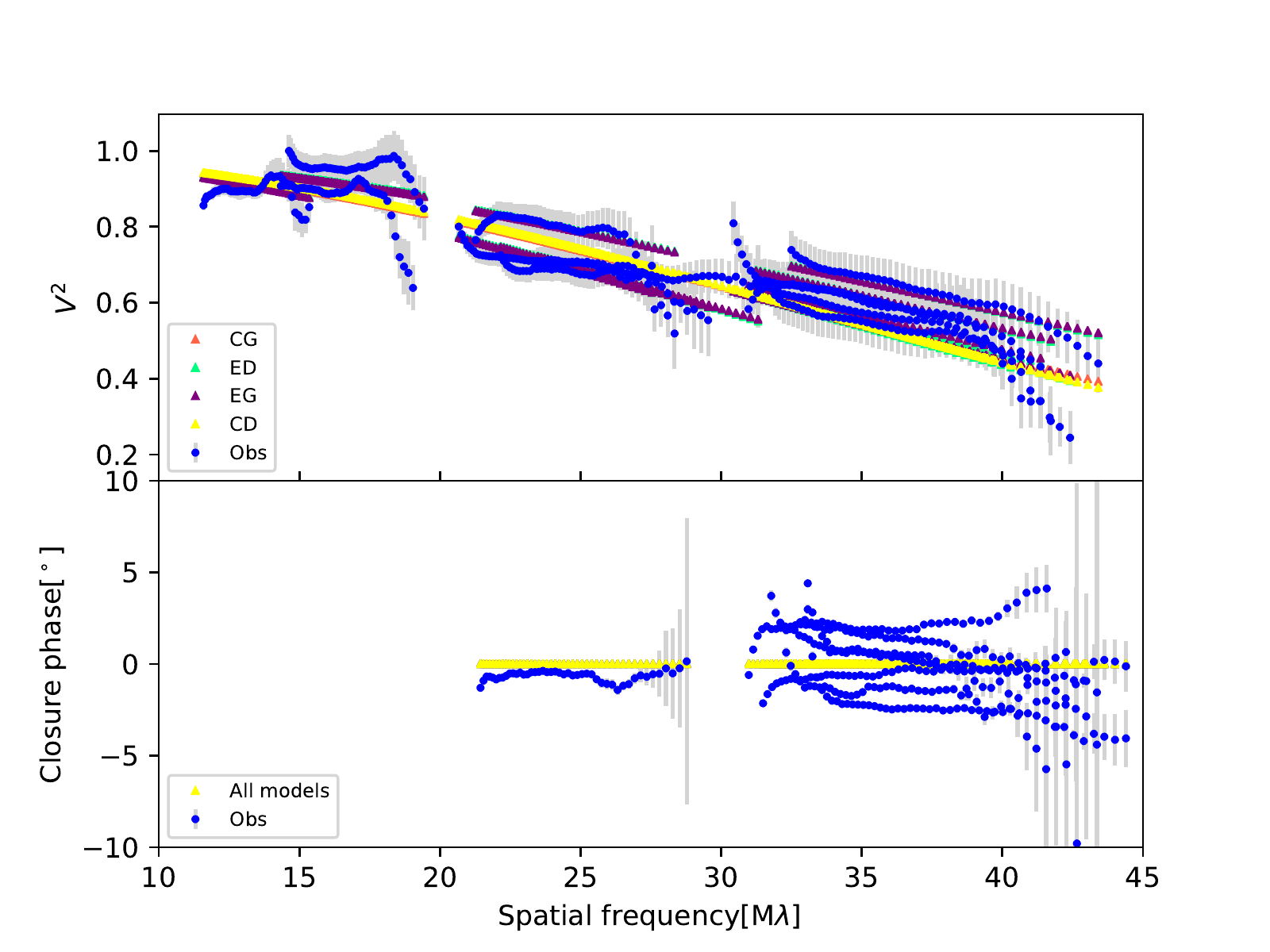}
    \caption{Squared visibilities and closure phases as functions of spatial frequency for A1 models (cf. Table\,\ref{table:Table 3}). Triangles represent the models while the blue points represent the $L$ band MATISSE observables.}
    \label{Fig-A1}
\end{figure*}

We used three approaches to model the data. The resulting best-fit models are characterized in Table\,\ref{table:Table 3}, where we provide formal fit errors from LITpro which are largely underestimated. The first approach, hereafter called A1, involved fitting a single uniform disk or a Gaussian, both either circular or elliptical, whose central position was fixed at the center of the FoV. The best fits yield the diameter in the case of disks, or FWHM for Gaussian models. All A1 results, shown in Fig.\,\ref{Fig-A1}, imply a very small source of less than about 3\,mas, comparable to our resolution at longest baselines. Non-circular models consistently imply a small flattening of the structure (a stretch ratio of 1.4 $\pm$ 0.1) with the longer axis at a PA of $\approx$--40\degr. The elliptical models are only slightly better than the circular ones. The A1 models produce zero closure phases and thus do not reproduce the observed deviations of up to $\approx\pm$5\degr\ at longer baselines.

\begin{figure*}
    \centering\sidecaption
    \includegraphics[width=6cm]{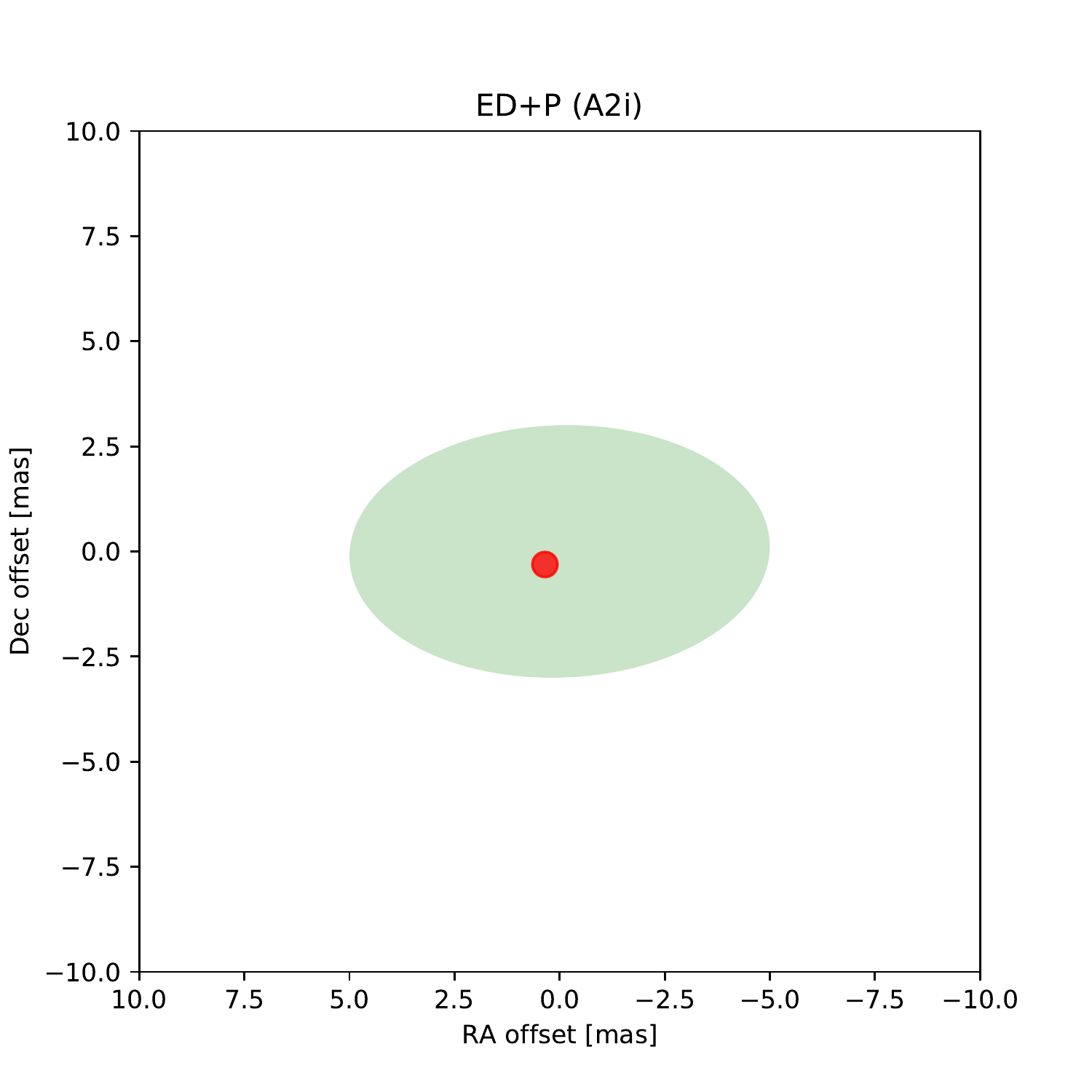}
    \includegraphics[width=6cm]{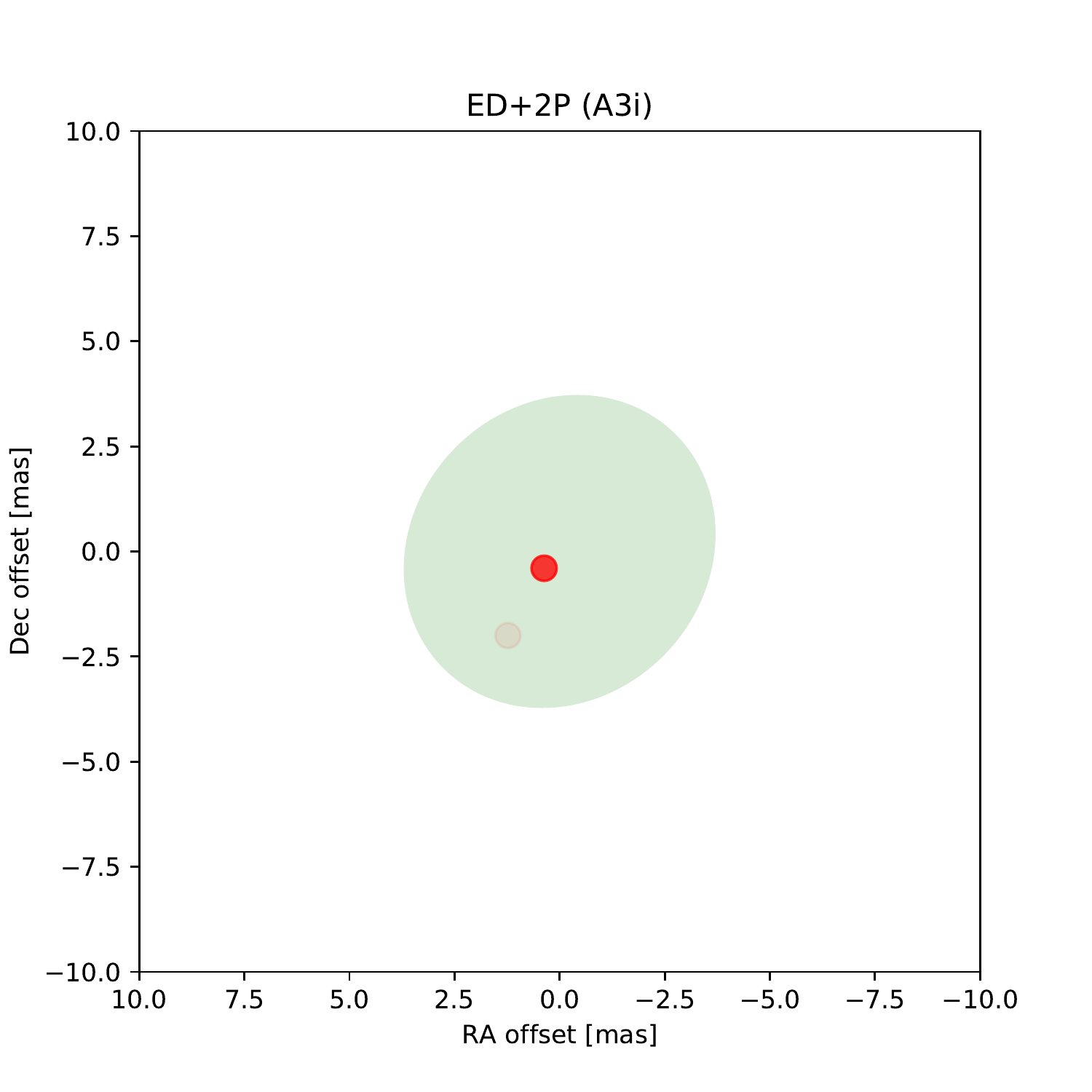}
    \caption{Graphical representation of a sample of simple geometrical models. Left: The A2i model for elliptical disk and a single point source. Right: The A3i model with and elliptical disk and two point sources (red circles). The transparency of the points corresponds to their relative fluxes.}
    \label{Fig-models}
\end{figure*}

In the second approach, A2, we attempted to represent the data by an extended component and a single point source. This was done in two ways, A2i and A2ii. In A2i, we simultaneously fitted the parameters for both  components (except for the position of the extended component, which was fixed at the FoV center). In A2ii, we first fitted the extended component, fixed the obtained parameters, added the point source, and searched for its best-fit parameters (the relative flux level, and the relative $x$ and $y$ position). The best-fit results are shown in Figs.\,\ref{Fig-A2i} and \ref{Fig-A2ii}. The addition of a point source at an offset of 0.43--0.47\,mas in A2i or 0.95--1.49\,mas in A2ii north-west from the center introduces a slight asymmetry to the models, which is consistent with the observed non-zero closure phases. In A2i models, the flux of the point source dominates over the extended component. Also in A2i, the extended structure is always a few times larger than in A1. The eccentricity of the elliptical A2i models is increased with respect to A1 and the longer axis is oriented latitudinally, as illustrated graphically in Fig.\,\ref{Fig-models}. The A2ii approach favors models with a dominance of the extended component over the point source and with the point source at a larger offset than in A2i. The offsets are however very small compared to that of the companion of V838\,Mon, and thus could potentially represent inhomogenities in the circumstellar dust.

Models A2 are probably already too complex to be well constrained by the limited observations in hand. However, to investigate whether there is a chance that part of the observed flux comes from the dust-enshrouded companion, we extended the A2 models by introducing a second point source. In the first form of this approach (A3i) we simultaneously fit for all the parameters of the three components (i.e. the extended source and both point sources, while the position of the extended source is fixed in the field center). A sample best-fit model of A3i is sketched in Fig.\,\ref{Fig-models}. In A3ii, we first fitted the data only with a disk or a Gaussian and obtained all the parameters of this component (as in A1). Then we fixed its parameters, except for the flux level and location. We next added the first point source and ran the fitting procedure to obtain relative fluxes of both components and the position of the point source. We next added a second point source and searched for its best fit position and relative fluxes of the three components. Results for the A3 models are shown in Table\,\ref{table:Table 3} and Figs.\,\ref{Fig-A3i} and \ref{Fig-A3ii}. For the best models with two point sources, one of the points has typically an insignificant flux, of a few \%, at or below the sensitivity limit of the observations. All point sources in the A3 models are located within $\approx$5\,mas from each other, and therefore none can represent the distant  B type companion known from ALMA maps (at a separation of 38\,mas).  

While models A3i have the lowest $\chi^2_r$ in Table\,\ref{table:Table 3}, we do not consider them to be the best representation of the observed source. Their good formal fit is a consequence of many free model parameters. The simpler models A1 and A2 are more robust and informative about the source structure given the limited data. Overall, V838\,Mon observed in $L$ band is a slightly elongated source with an ellipticity of about 1.4 and the major axis at a PA of $\approx$--40\degr. Its longer extent is about 3.3\,mas (or FWHM of about 2.0\,mas). From A2 models we conclude that the dominant emission component may be located slightly off center or that there is a feature northeast from the FoV center that introduces a slight asymmetry in the system. The elliptical models of A2i seem to suggest an orientation of the extended component that differs from the value of --40\degr\ mentioned earlier. However, given that the point source is located along a PA of --48\degr, these models also suggest the same overall geometry as the non-circular models used in A1.  

\subsection{Image reconstruction}\label{sec-reconstr}

\begin{figure}
    \centering
    \includegraphics[trim=0 0 40 20, clip, width=\columnwidth]{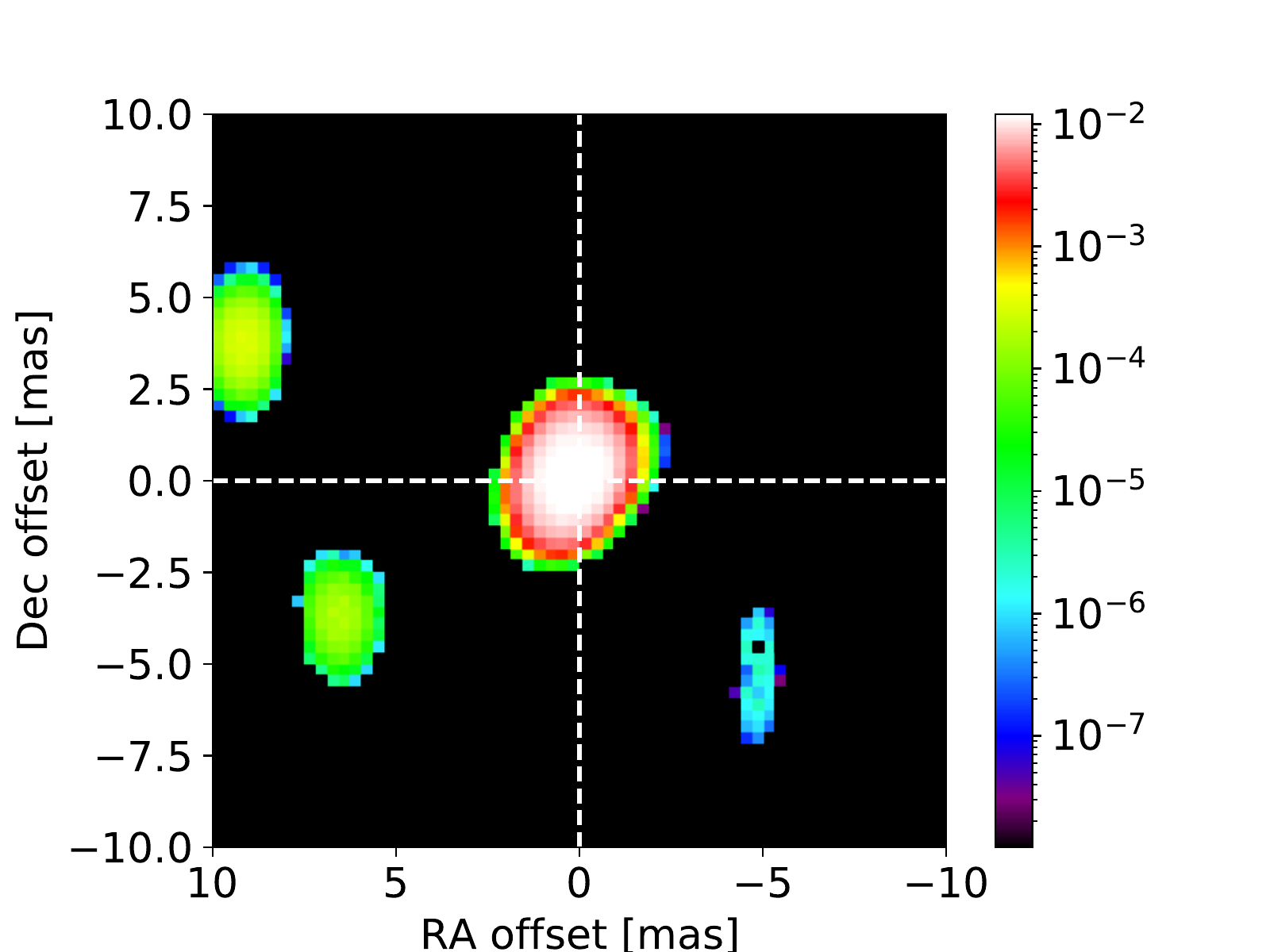}
    \caption{Reconstructed image of V838 Mon in the $L$ band range 2.95--4\,$\mu$m. The low-intensity features are imaging artifacts. The colors represent brightness in arbitrary units. }
    \label{fig-v838image}
\end{figure}

Although comprehensive imaging was not the goal of this study due to the sparse data, we did reconstruct a single image of V838\,Mon in the $L$ band. We made use of  OImaging\footnote{\url{https://www.jmmc.fr/english/tools/data-analysis/oimaging/}} software. A model-independent image was reconstructed with the Weak-phase Interferometric Sample Alternating Reconstruction Device (WISARD), an algorithm described in \cite{2008JOSAA..26..108M}. We fixed the FoV to 20\,mas and set the image size to 64$\times$64 pixels. We used the default regularization LIL2, along with the default values for the convergence threshold and regularization scale factors. The maximum number of iterations was set to 50. The image was reconstructed using the squared visibilities and closure phases. The resulting image is shown in Fig.\,\ref{fig-v838image}. The centroid of the main emission component is located slightly (<1 mas) off center. Several weaker structures can be seen in Fig.\,\ref{fig-v838image}, but these are much weaker than a limit set by the dynamic range of the data and thus can be treated as artifacts.    

\subsection{Testing ALMA model}
\begin{figure}
    \centering
    \includegraphics[width=\columnwidth]{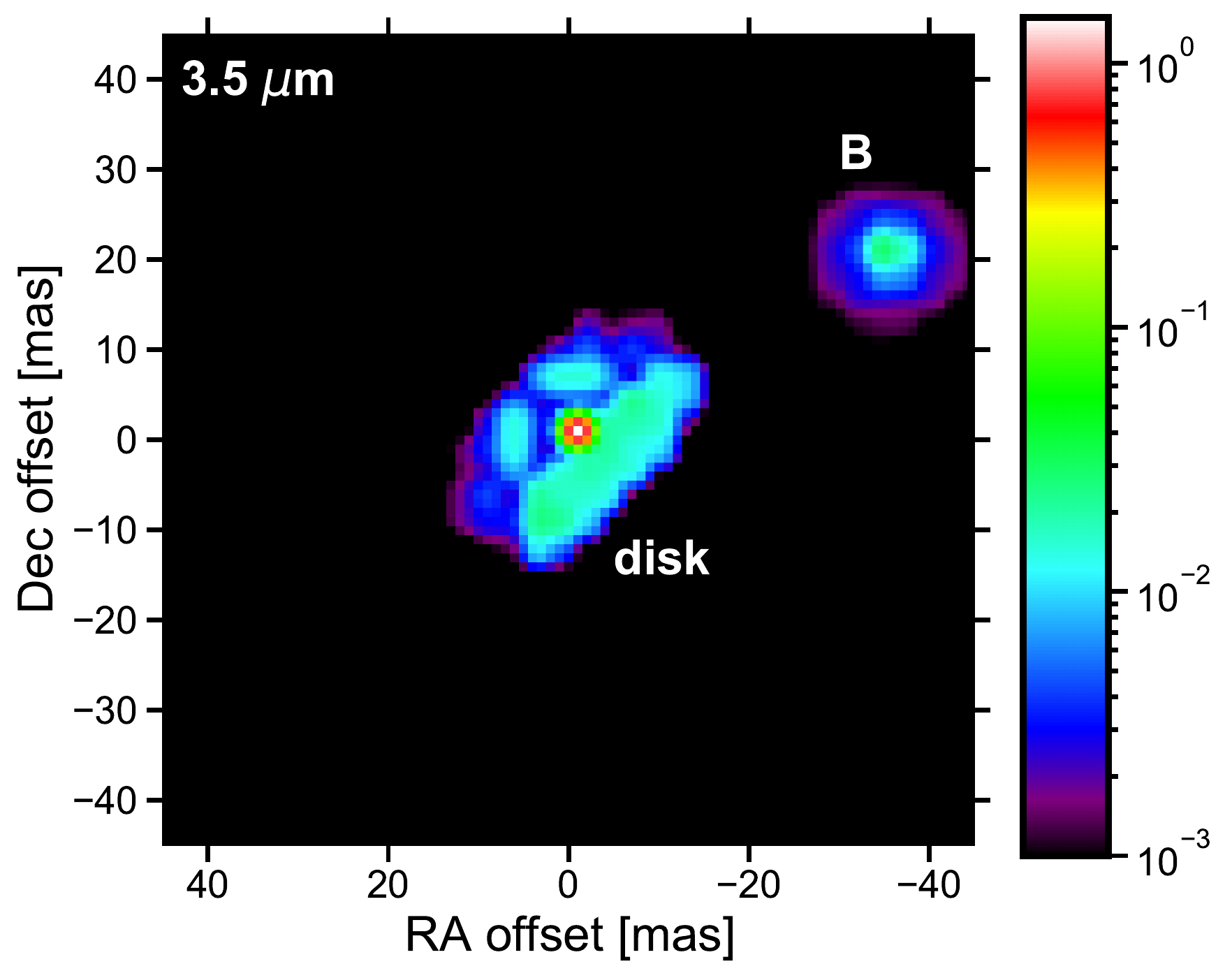}
    \caption{Image of the V838\,Mon model at 3.5\,$\mu$m based on the SED and ALMA observations. The component seen northwest is related to the B-type companion. For the presentation purposes, the image was smoothed to a resolution of 2\,mas. Colors represent brightness in units of Jy per beam.}
    \label{fig-almamodel}
\end{figure}

Based on SED and high-resolution maps of continuum emission at 1.3\,mm, \cite{alma} constructed a 3D model of dust distribution of the entire remnant of V838\,Mon. This was performed for epochs contemporary to our MATISSE observations. Their ALMA data revealed the presence of a mm-wave continuum source, component M, of a size of 17.6$\times$7.6\,mas, which partially aligns with the simple elongated models considered above for the $L$ band data. Component M was arbitrarily explained by a disk seen at moderate inclinations ($\sim$20\degr) but ALMA resolution of 20\,mas was still too poor to put good constraints on the form or nature of the dusty structure. Nevertheless, the MATISSE data allow us to test the model predictions for the $L$ band. Using a ray-tracing algorithm of the RADMC-3D software \citep{radmc3d}, we generated a sky image of the model at 3.5\,$\mu$m which is shown in Fig.\,\ref{fig-almamodel}. It is dominated by a point source (<2\,mas) surrounded by disk material, whose emission is seen only at a level of only a few \% of that of the point source. We next generated squared visibilities and closure phases for this image with baselines consistent with the MATISSE observations. Their match to the MATISSE observations is however very poor. The model fluxes are too strongly concentrated in the point source, producing almost flat values of $V^2$ as a function of spatial frequency. The brightness of the circumstellar material in the immediate vicinity of V838\,Mon must be higher than assumed in the model of \cite{alma}. 

\section{Summary and discussion}
The modeling attempts in Sects.\,\ref{sec-litpro} and \ref{sec-reconstr} identified a slightly elongated source (ellipticity of about 1.4) of a size of about 3\,mas whose substructure remains unresolved. The deviation from a centro-symmetric configuration, as evident in closure phase measurements, may be the result of low-level inhomogeneities present in the dusty medium surrounding V838\,Mon. The MIR observations do not show signatures of the dust-enshrouded companion of V838\,Mon, which remains too faint to be observed in the $L$ band. 

\begin{figure}
    \centering
    \includegraphics[width=\columnwidth]{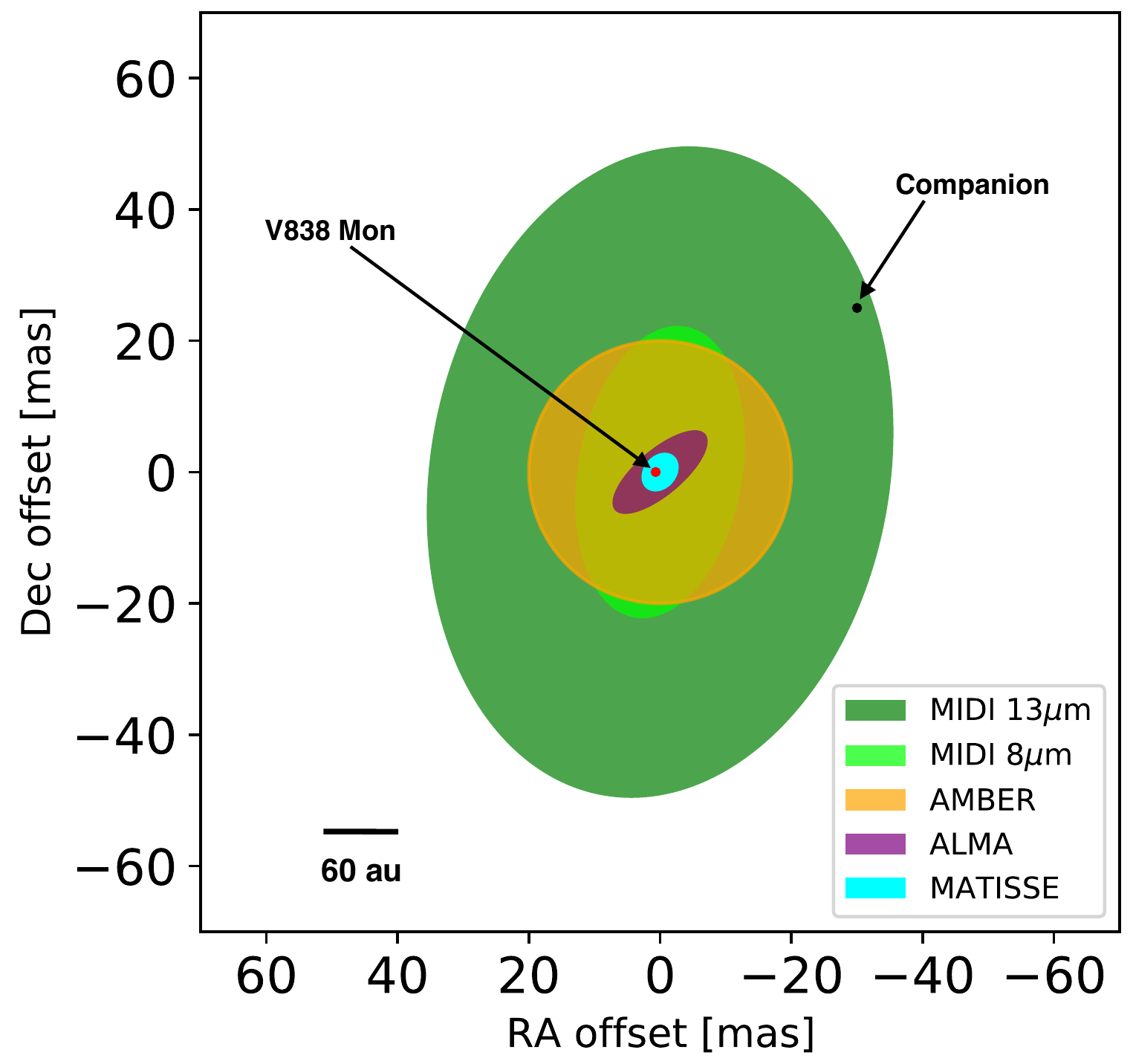}
    \caption{Sketch of the multi-wavelength structure of V838\,Mon based on observations that span from the MIR to the mm regime.} 
    \label{fig-v838sketch}
\end{figure}

Most of the considered models of the MATISSE data lead to configurations elongated at a PA of --40\degr. This angle is close to angles of elongation found for structures immediately surrounding V838\,Mon and probed at a wide range of wavelengths (1.7\,$\mu$m--1.3\,mm). Their locations and orientations are schematically represented in Fig.\,\ref{fig-v838sketch}. Using MIDI in $N$ band, \cite{2014A&A...569L...3C} found an extended dusty envelope with a size that increases with wavelength. Gaussian fits yielded FWHM of 25 mas at 8 $\mu$m to 70\,mas at 13 $\mu$m and a position angle of --10\degr$\pm$30\degr. This PA is consistent with our result for the $L$-band source. Their $H$ and $K$ band data from AMBER suggested further that the $N$-band structure has a corresponding near-IR feature of a size larger than about 20\,mas (but of unconstrained geometry and orientation). At the other end of the spectrum, at mm wavelengths, \cite{alma} found near V838\,Mon an elongated dusty structure extending 18$\times$8\,mas and elongated at a PA of --46\degr$\pm$4\degr, again within the uncertainties consistent with the orientation of the $L$ band source. This close correspondence of orientations across the spectrum can be interpreted as different manifestations of the same physical structure, perhaps a disk or torus seen at a moderate inclination. However, the sparse infrared visibilities collected so far do not allow us to make any firm conclusions. In particular, a typical circumstellar disk with a dust temperature dropping with distance should appear most extended at longest wavelengths, which does not seem to be the case here. 
ALMA identified other dusty components within the remnant, some of which partially overlap with the outer structure revealed by MIDI. Since the analysis of \cite{2014A&A...569L...3C} was limited to simple geometric models, the comparison to other wavelengths may be misleading. Only full interferometric imaging at NIR and MIR wavelengths can demonstrate true associations across the spectrum and reveal the full architecture of the post-merger system.

Since the NIR and mm interferometric data on the extended dusty component span from 2011 to 2020 and all indicate virtually the same orientation, the probed structure within about 40\,mas appears to have stayed relatively stable for over a decade, if not longer. Interestingly, polarimetric observations of V838\,Mon in the optical showed significant linear polarization whose characteristics may suggest a relation to the currently observed feature. In February 2002, when the object was still in outburst, \cite{wisniewski2003b} found that the $R$ band light was polarized along a PA of 151\degr\ or --29\degr, close to the current orientation of the $L$-band structure. In October 2002, that is, in the decline phase and when dust formation might have been taking place, a 90\degr\ flip in the intrinsic polarization angle was observed in the optical \citep{wisniewski2003a}. There is no clear explanation of the flip or even of the origin of the polarized light in the different epochs, but these early observations further suggest a persistent geometry in the immediate surroundings of V838\,Mon for almost two decades. That geometry is missing in molecular emission probed by ALMA \citep{alma}. 

It is most tempting to assume that the MIR and mm interferometric data are signatures of a circumstellar disk surrounding the recently coalesced binary. There is plenty of evidence of circumstellar disks in nearly all known or suspected post-merger systems, including V1309\,Sco \citep{v1309a,v1309b}, V4332\,Sgr \citep{v4332}, and CK\,Vul \citep{ckvul}. The disks are also predicted to form by simulations of common-envelope or merging systems and often represent material lost by the binary through the L$_2$ point into the orbital plane before the final plunge-in phase \citep[e.g.,][]{pejcha2016,macleod2018,macleod2020}. If confirmed, the observed feature would then define the orientation of the orbit of the binary before its catastrophic collision. As the flattened structure potentially contains important information regarding the processes that cause stellar collisions, further IR interferometric studies are vital and should aim to fully map the complex system of V838\,Mon using model-independent imaging techniques.

\begin{acknowledgements}
T. K. and M. Z. M. acknowledge funding from grant no 2018/30/E/ST9/00398 from the Polish National Science Center. Based on observations made with ESO telescopes at Paranal observatory under program ID 0104.D-0101(C). This research has benefited from the help of SUV, the VLTI user support service of the Jean-Marie Mariotti Center (\url{http://www.jmmc.fr/suv.htm}). This research has also made use of the JMMC's  Searchcal, LITpro, OIFitsExplorer, and Aspro services.
\end{acknowledgements}

\bibliographystyle{aa}
\bibliography{export-bibtex.bib}

\begin{landscape}
\begin{table}[ht]
\caption{Fitted models and their parameters.}
\centering
\begin{tabular}{l c c cc  c c c cccc}
\hline\hline
Model & $\chi^2_{\rm r}$ & Size  & $PA$  & Stretch & EX1 flux  & P1 flux  & P2 flux & $x$1 & $y$1 & $x$2 & $y$2  \\ 
& & [mas]&[\degr]& ratio &[\%]&[\%]&[\%]&[mas]&[mas]&[mas]&[mas] \\
\hline
CD(A1)  &50.9  &2.80  $\pm$  0.08 &- &- &100\tablefootmark{a} &- &- &- &- &- &- \\
CG(A1)  &50.9 &1.69  $\pm$  0.05 &- &- &100\tablefootmark{a} &- &- &- &- &- &- \\
ED(A1)  &50.2  &3.28  $\pm$  0.18 &--40  $\pm$  6 &1.4  $\pm$  0.1 &100\tablefootmark{a} &- &- &- &- &- &-\\
EG(A1)  &50.2 &1.96  $\pm$  0.10 &--40  $\pm$  6  &1.4  $\pm$  0.1 &100\tablefootmark{a} &- &- &- &- &- &-\\[5pt]
CD+P(A2i)  &18.3 &8.38   $\pm$  0.18 &- &- &18  $\pm$  3 &82  $\pm$  12 &- &0.30 $\pm$ 0.02 &--0.32 $\pm$ 0.02 &- &-\\ 
CG+P(A2i) &18.5 &4.63  $\pm$  0.16 &- &- &23  $\pm$  3 &77  $\pm$  11 &- &0.33 $\pm$ 0.02 &--0.36 $\pm$ 0.02 &- &-\\
ED+P(A2i)  &9.4 &9.17  $\pm$  0.16 &--88  $\pm$  1 &1.6  $\pm$  0.1 &21  $\pm$  2 &79  $\pm$  8 &- &0.32 $\pm$ 0.02 &--0.28 $\pm$ 0.01 &- &-\\
EG+P(A2i) &9.5 &5.12  $\pm$  0.11& --90  $\pm$  1& 1.6  $\pm$  0.1 & 24  $\pm$  3 &76 $\pm$  8&- &0.35 $\pm$ 0.02 &--0.31 $\pm$ 0.02 &- &- \\[5pt]
CD+P(A2ii) &14.7 &2.80 $\pm$  0.08\tablefootmark{b} &- &-&83  $\pm$  11 &17  $\pm$  2 &- &0.72  $\pm$  0.03 &--1.31  $\pm$  0.02 &- &- \\               
CG+P(A2ii) &14.3 &1.69  $\pm$  0.05\tablefootmark{b}&-&-&82  $\pm$  10& 18  $\pm$  2 &- &0.72  $\pm$  0.02 &--1.31 $\pm$ 0.02 &- &- \\
ED+P(A2ii) &19.8 &3.28  $\pm$  0.18\tablefootmark{b}&--40  $\pm$ 6\tablefootmark{b} & 1.4  $\pm$  0.1\tablefootmark{b}& 57  $\pm$  8& 43 $\pm$ 6&- &0.74 $\pm$ 0.02 &--0.60 $\pm$ 0.03 &- &-\\
EG+P(A2ii) &19.5 &1.96  $\pm$  0.10\tablefootmark{b}& --40  $\pm$  6\tablefootmark{b}& 1.4  $\pm$  0.1\tablefootmark{b}& 56  $\pm$  8&44  $\pm$  7&- &0.76 $\pm$ 0.02 &--0.63 $\pm$ 0.03 &- &-\\[5pt] 
CD+2P(A3i)  &8.4 &7.84  $\pm$  0.31 &- &- &19  $\pm$  2 &80  $\pm$  2&1.0  $\pm$  0.1 &0.12 $\pm$ 0.02 &--0.22 $\pm$ 0.12 &--3.62 $\pm$ 0.20 &0.83 $\pm$ 0.12\\ 
CG +2P(A3i)  &8.5 &4.07  $\pm$  0.22 &- &- & 24  $\pm$  3 &75  $\pm$  7 & 1  $\pm$  0.1 &0.12 $\pm$ 0.02&--0.23 $\pm$ 0.02&--3.74 $\pm$ 0.17&0.85 $\pm$ 0.11 \\ 
ED+2P(A3i) &6.8 &8.62  $\pm$  0.30 & --44  $\pm$  18 &1.1  $\pm$  0.1 & 16  $\pm$  2  & 7  $\pm$  4 & 77  $\pm$  7 &1.23 $\pm$ 0.19 &--2.01 $\pm$ 0.31 &0.37 $\pm$ 0.03 &--0.40 $\pm$ 0.06\\
EG+2P(A3i) &7.1 &4.52  $\pm$  0.22&--43  $\pm$  28& 1.1  $\pm$  0.1& 19.0  $\pm$  0.1& 8.0  $\pm$  0.1& 73  $\pm$  7 &1.23 $\pm$ 0.21&--2.00 $\pm$ 0.40&0.40 $\pm$ 0.04&--0.42 $\pm$ 0.07\\ [1ex] 
CD+2P(A3ii) &10.5  &2.8  $\pm$  0.1\tablefootmark{b} &- &- &68 $\pm$  7  &14  $\pm$  11 &18  $\pm$  10 &0.73 $\pm$ 0.06 &--1.43 $\pm$ 0.36 &0.66 $\pm$ 0.05 &--0.41 $\pm$ 0.43\\
CG+2P(A3ii)&10.1 &1.69  $\pm$  0.05\tablefootmark{b} &-&-&68  $\pm$  7& 14  $\pm$  11& 18  $\pm$ 11 &0.74 $\pm$ 0.06 &--1.43 $\pm$ 0.35 &0.69 $\pm$ 0.05 &--0.43 $\pm$ 0.43\\
ED+2P(A3ii) &11.1 &3.28  $\pm$  0.18\tablefootmark{b} &--40  $\pm$  6\tablefootmark{b} &1.4  $\pm$  0.1\tablefootmark{b} &58  $\pm$  8  &6  $\pm$  7 &36  $\pm$  7&1.44 $\pm$ 0.39 &--2.14 $\pm$ 0.73 &0.66 $\pm$ 0.10 &--0.70 $\pm$ 0.20 \\
EG+2P(A3ii)  &10.9 &1.96  $\pm$  0.10\tablefootmark{b} &--40  $\pm$  6\tablefootmark{b} &1.4  $\pm$  0.1\tablefootmark{b} &57  $\pm$  7  &5  $\pm$  6 &38  $\pm$  6 &1.50 $\pm$ 0.41 &--2.23 $\pm$ 0.76 &0.69 $\pm$ 0.09 &--0.75 $\pm$ 0.18\\[5pt] 
\hline
\end{tabular}
\label{table:Table 3}
\tablefoot{The models used to represent the $L$ band data from MATISSE. The parameters are PA of the major axis, percentage model flux of the non-point source component (EX1), percentage flux of the first point source (P1), and percentage flux of the second point source (P2); $x$1, $y$1, $x$2, and $y$2 are the offsets of the point sources relative to the center of the FoV. The basic models are a point source (P), a uniform circular or elliptical disk (CD or ED, respectively), a circular or elliptical Gaussian (CG or EG, respectively). In the case of Gaussians, size refers to the FWHM of the major axis, while for disks it is the angular diameter along the major axis. The stretch ratio is between that of the major and minor axis size. \tablefoottext{a} Flux value was fixed from the start. \tablefoottext{b} The parameters of the extended component were fixed after they were first obtained.}

\end{table}
\end{landscape}

\begin{appendix}
\section{More geometric models} 
\begin{figure}
    \centering
    \includegraphics[trim=0 0 20 30, clip, width=\columnwidth]{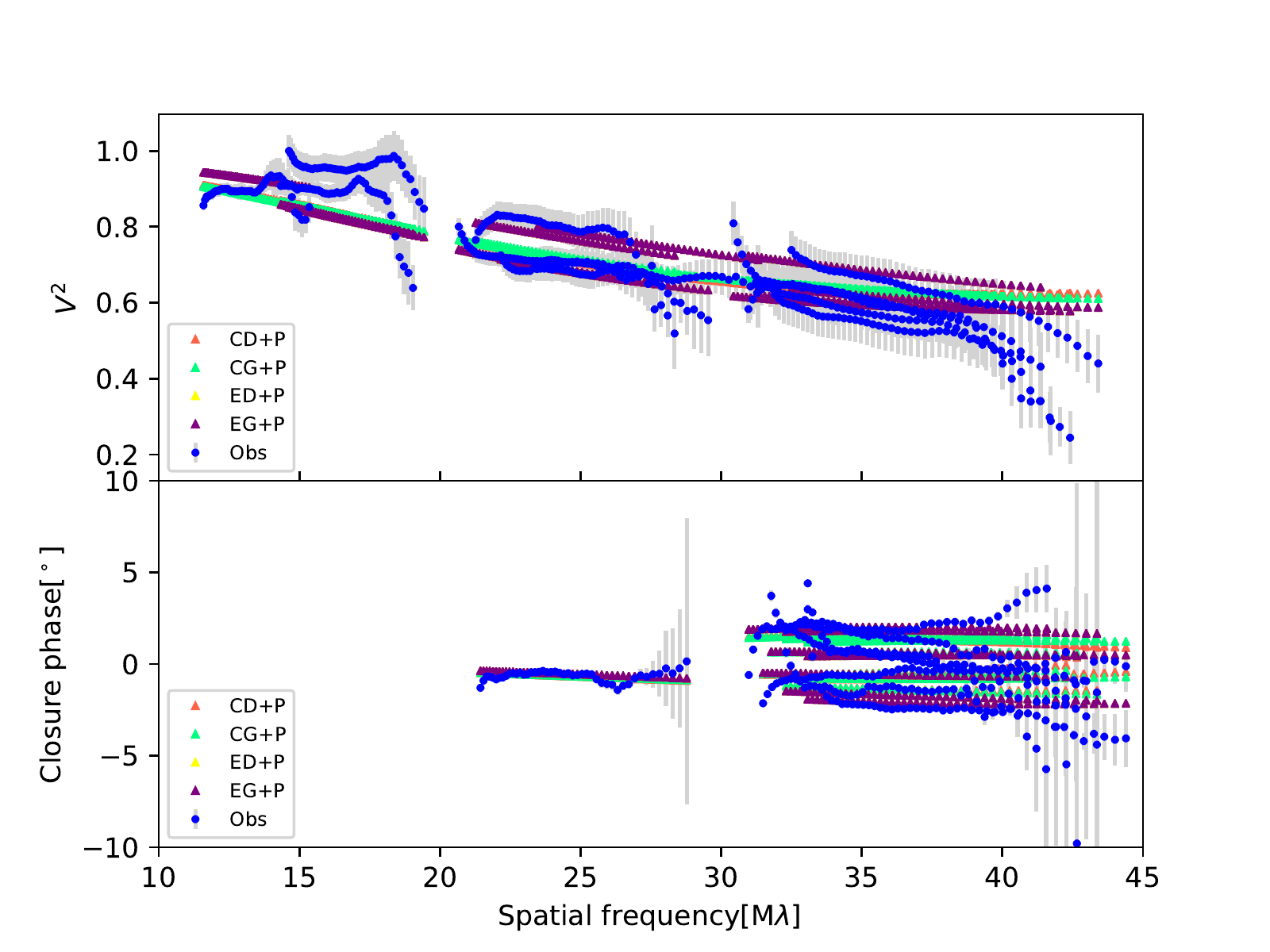}
    \caption{Same as Fig.\,\ref{Fig-A1} but for A2i models.}
    \label{Fig-A2i}
\end{figure}

\begin{figure}
    \centering
    \includegraphics[trim=0 0 20 30, clip, width=\columnwidth]{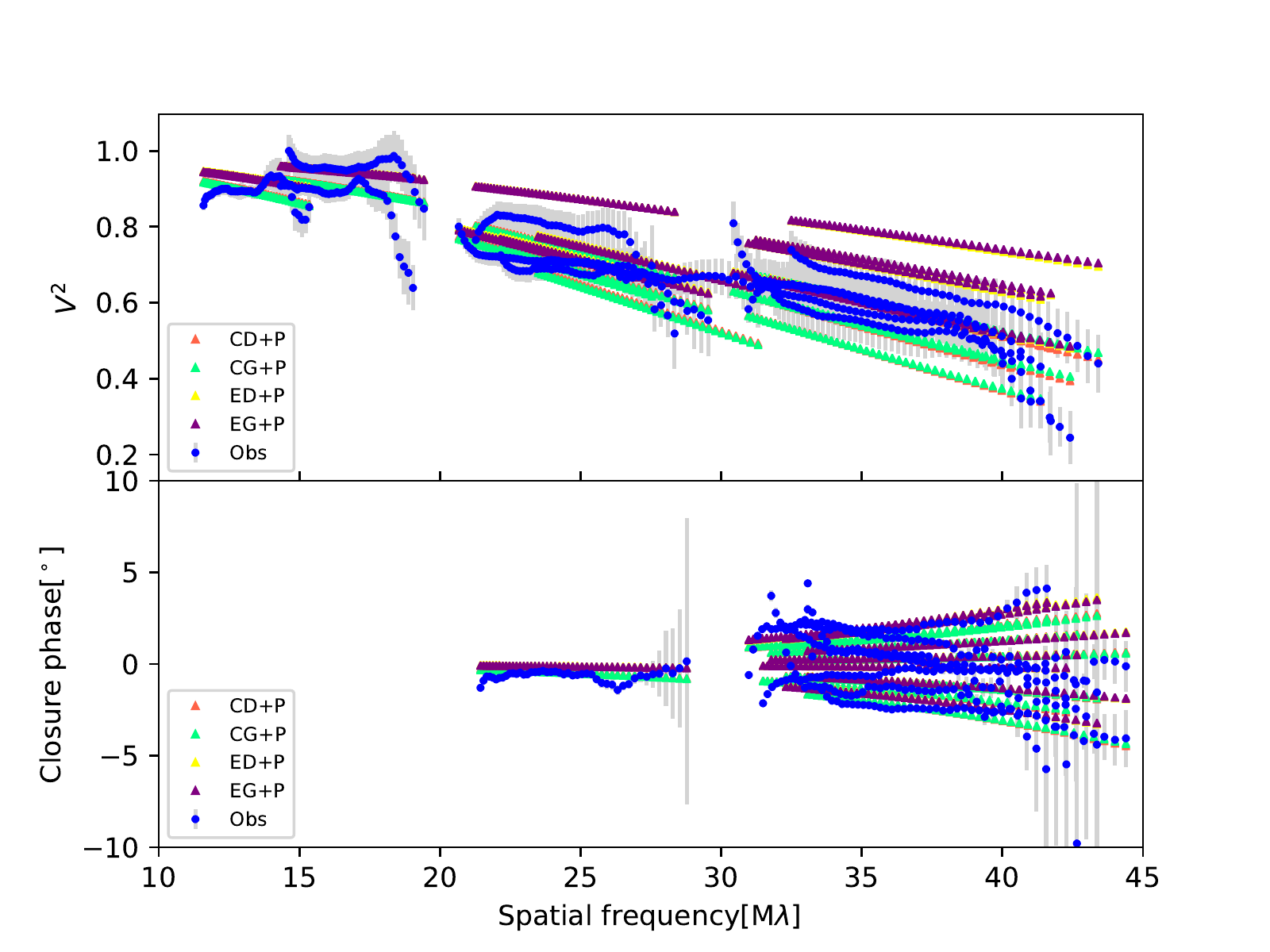}
    \caption{Same as Fig.\,\ref{Fig-A1} but for A2ii models.}
    \label{Fig-A2ii}
\end{figure}

\begin{figure}
    \centering
    \includegraphics[trim=0 0 20 30, clip, width=1\columnwidth]{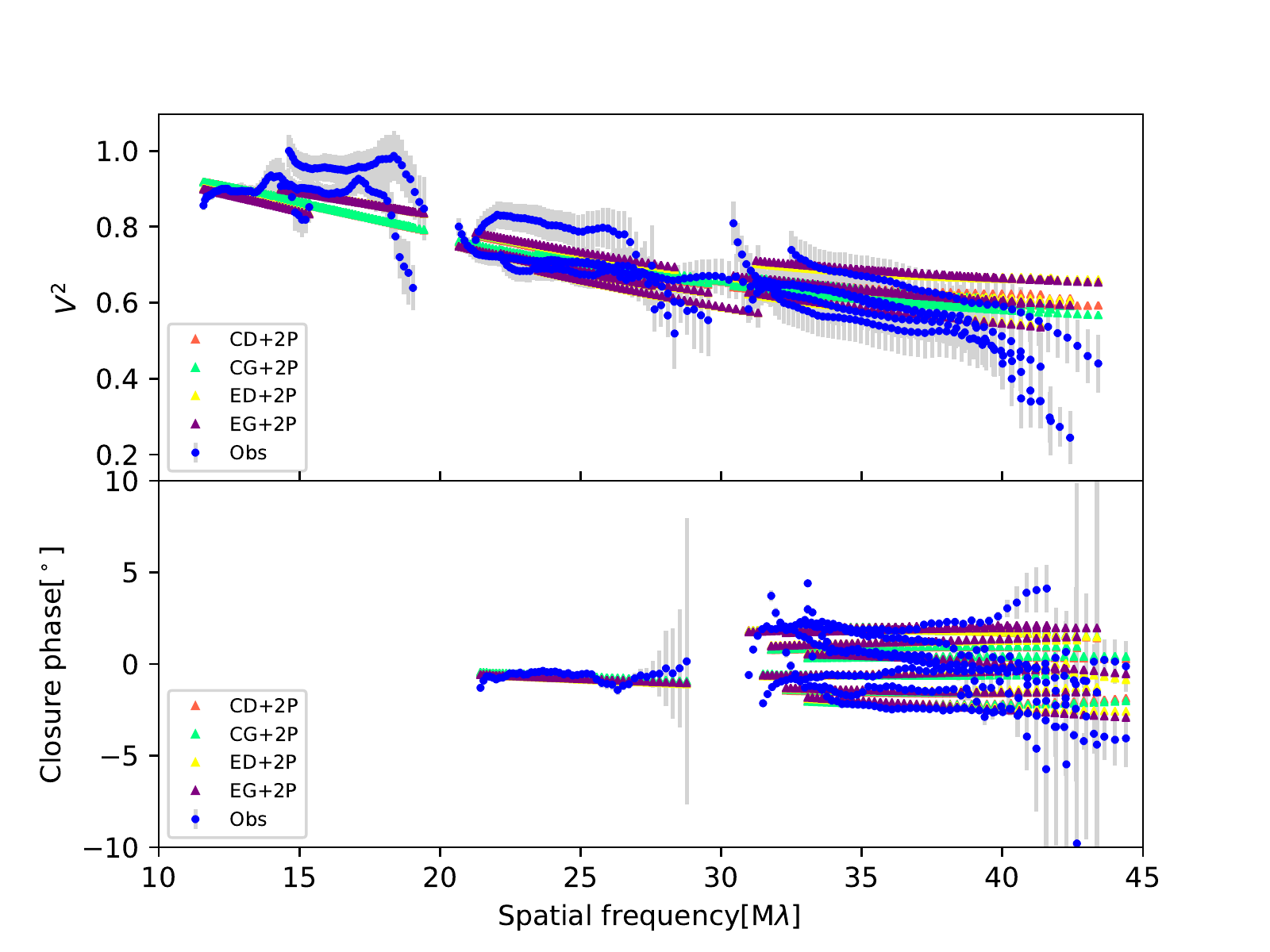}
    \caption{Same as Fig.\,\ref{Fig-A1} but for A3i models.}
    \label{Fig-A3i}
\end{figure}

\begin{figure}
    \centering
    \includegraphics[trim=0 0 20 30, clip, width=\columnwidth]{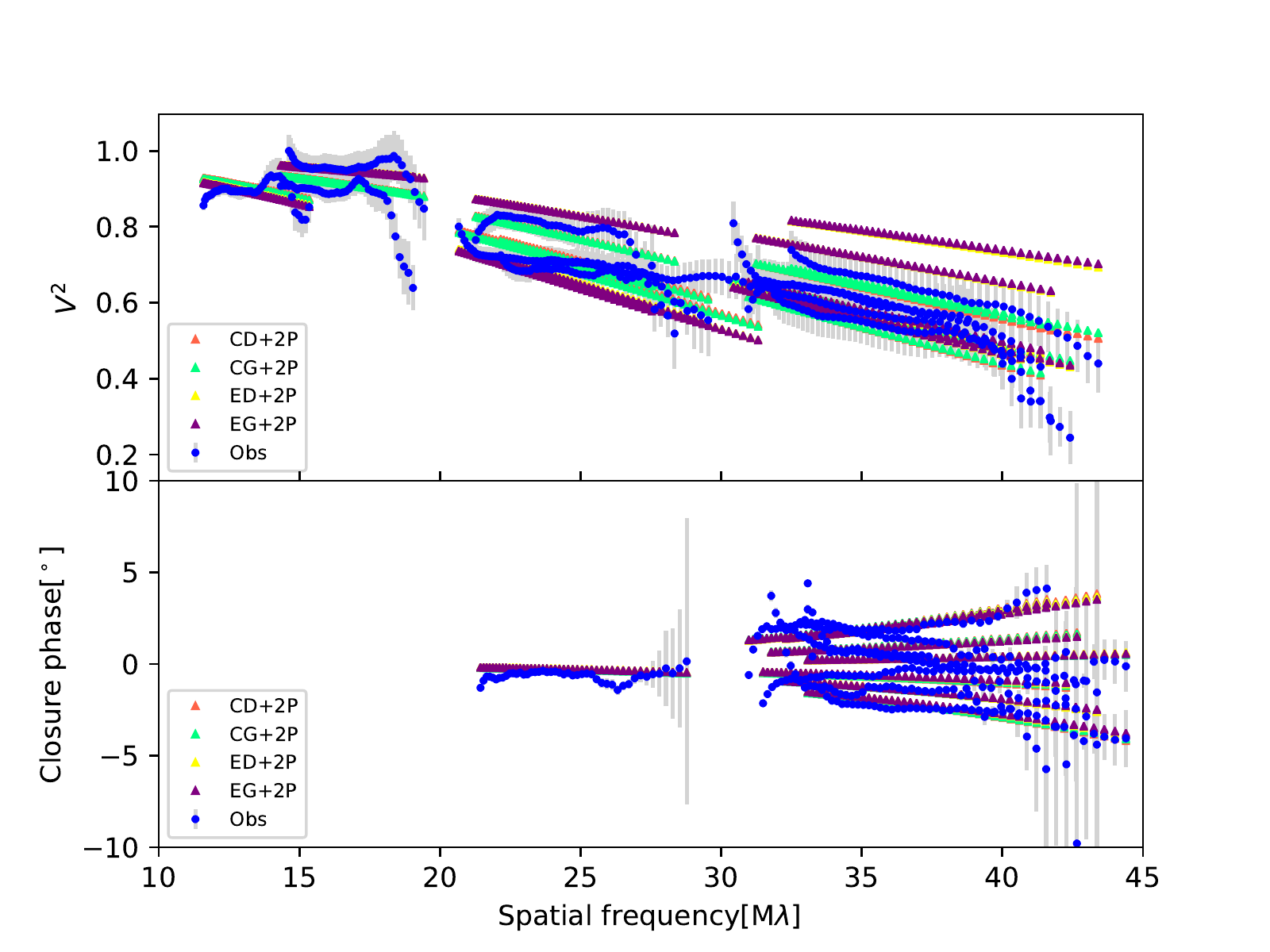}
    \caption{Same as Fig.\,\ref{Fig-A1} but for A3ii models.}
    \label{Fig-A3ii}
\end{figure}

Figures  \ref{Fig-A2i} to \ref{Fig-A3ii} show comparisons of the observed MATISSE quantities to best-fit models of simple geometrical configurations considered in Sect.\,\ref{sec-litpro}.


\end{appendix}

\end{document}